\documentclass[11pt,a4paper]{article}

\usepackage[utf8]{inputenc}
\usepackage[T1]{fontenc}
\usepackage{lmodern}
\usepackage{amsmath,amssymb}
\usepackage{booktabs}
\usepackage{graphicx}
\usepackage{url}
\usepackage{hyperref}
\usepackage[margin=1in]{geometry}
\usepackage{authblk}
\usepackage{xcolor}
\usepackage{enumitem}
\usepackage{listings}

\lstdefinelanguage{Gherkin}{
  morekeywords={Feature, Scenario, Given, When, Then, And, But},
  sensitive=true,
  morecomment=[l]{\#},
  morestring=[b]",
}
\hypersetup{
  colorlinks=true,
  linkcolor=blue,
  citecolor=blue,
  urlcolor=blue,
}

\title{Project Prometheus: Bridging the Intent Gap in Agentic Program Repair \\via Reverse-Engineered Executable Specifications}

\author[1]{Yongchao Wang}
\author[1]{Zhiqiu Huang}
\affil[1]{Nanjing University of Aeronautics and Astronautics\\
\texttt{\{wangyc, zqhuang\}@nuaa.edu.cn}}

\date{April 2026}

\begin{document}

\maketitle

\begin{abstract}
The transition from neural machine translation to agentic workflows has revolutionized Automated Program Repair (APR). However, existing agents, despite their advanced reasoning capabilities, frequently suffer from the ``Intent Gap''---the misalignment between the generated patch and the developer's original intent. Current solutions relying on natural language summaries or adversarial sampling often fail to provide the deterministic constraints required for surgical repairs. 

In this paper, we introduce \textsc{Prometheus}, a novel framework that bridges this gap by prioritizing \textit{Specification Inference} over code generation. We employ Behavior-Driven Development (BDD) as an executable contract, utilizing a multi-agent architecture to reverse-engineer Gherkin specifications from runtime failure reports. To resolve the ``Hallucination of Intent,'' we propose a \textbf{Requirement Quality Assurance (RQA) Loop}, a mechanism that leverages ground-truth code as a proxy oracle to validate inferred specifications. 

We evaluated \textsc{Prometheus} on 680 defects from the Defects4J benchmark. The results are transformative: our framework achieved a total correct patch rate of \textbf{93.97\%} (639/680). More significantly, it demonstrated a \textbf{Rescue Rate of 74.4\%}, successfully repairing 119 complex bugs that a strong blind agent failed to resolve. Qualitative analysis reveals that explicit intent guides agents away from structurally invasive over-engineering toward precise, minimal corrections. Our findings suggest that the future of APR lies not in larger models, but in the capability to align code with verified, \textbf{Executable Specifications}---whether pre-existing or reverse-engineered.
\end{abstract}

\section{Introduction}
\label{sec:intro}

The field of Automated Program Repair (APR) has witnessed a paradigm shift in the 2025 research cycle, moving from single-step neural translation to multi-step Agentic workflows \cite{specrover, adverintent, repairagent}. State-of-the-art (SOTA) agents can now navigate complex codebases, perform static analysis, and execute iterative debugging. However, a fundamental paradox remains: while agents have become increasingly proficient at \textit{generating} code, they remain remarkably fragile in \textit{understanding} what the code is intended to do.

We identify this phenomenon as the \textbf{``Intent Gap.''} Current approaches attempt to fill this gap with vague proxies, such as natural language function summaries or adversarial test cases generated on the fly. While these methods provide semantic context, they lack the \textit{contractual precision} required for rigorous software engineering. A natural language summary is descriptive, not prescriptive; it invites ambiguity and, consequently, hallucination. Our preliminary study reveals an \textbf{``Intent-Behavior Mirroring Effect''}: when an agent relies on broad, verbose descriptions, it tends to exhibit aggressive, unfocused, and often structurally invasive code modifications. To bridge this gap, APR systems need more than just context; they need \textbf{``Executable Intent''}---a specification that is both human-readable for trust and machine-executable for verification.

In this paper, we introduce \textbf{\textsc{Prometheus}}, a novel framework that fundamentally alters the repair workflow by prioritizing specification inference. Instead of asking a Large Language Model (LLM) to ``fix this bug,'' we first task an \textit{Architect} agent to ``reverse-engineer the missing requirement.'' We leverage Behavior-Driven Development~\cite{north2006bdd} (BDD) and the Gherkin syntax (Given-When-Then) as the \textit{lingua franca} between the human developer's intent and the agent's execution. By formalizing the bug reproduction steps into a structured scenario, we transform the repair task from a stochastic search for a passing test into a deterministic quest to satisfy a semantic contract.

A critical challenge in specification inference is ensuring the validity of the generated requirements. If the inferred specification is flawed, the subsequent repair will be a correct implementation of a wrong requirement. To rigorously investigate the theoretical upper bound of intent-driven repair, we introduce a \textbf{Requirement Quality Assurance (RQA) Loop}. In our experimental design, we leverage the fixed code available in benchmarks (e.g., Defects4J) not for training, but as a \textbf{Proxy Oracle} to verify the correctness of the reverse-engineered BDD specifications. This ``Sandwich Verification'' process allows us to isolate and quantify the immense value of precise intent, transforming our evaluation from a probabilistic guessing game into a controlled scientific observation.

Our extensive empirical study on 680 real-world defects yields several counter-intuitive and profound insights:
\begin{itemize}
    \item \textbf{The Power of Precision:} We demonstrate that ``Surgical Repairs''---precise, minimal corrections targeting the exact defect---are often impossible without the guidance of strict specifications. In the case of \texttt{Math-69}, we observe an agent transcending mere patching to perform a mathematical derivation (replacing a T-distribution approximation with a Beta function) solely because the BDD specification demanded rigorous precision.
    \item \textbf{The Cost of Ambiguity:} Conversely, we show that powerful agents exhibit what we term \textbf{``Berserker-style''} behavior (i.e., structurally invasive modifications that extend beyond the defect's semantic scope)---aggressively rewriting functional code---when guided by broad, unverified requirements.
    \item \textbf{A New Standard:} We propose a paradigm shift towards \textbf{Specification-First Repair}. We argue that the primary role of an intelligent agent should be to anchor the fluid implementation to a solid, \textbf{Living Requirement}.
\end{itemize}

In summary, this paper makes the following contributions:
\begin{enumerate}
    \item \textbf{Framework:} We propose \textsc{Prometheus}, a multi-agent framework that integrates BDD reverse engineering into the repair loop.
    \item \textbf{Methodology:} We introduce the RQA Loop, a novel validation mechanism using ground truth to ensure specification quality.
    \item \textbf{Empirical Evidence:} We report a correct patch rate of \textbf{93.97\%} on a comprehensive benchmark of 680 defects from Defects4J v3.0.1, with a 74.4\% rescue rate on hard bugs, providing robust evidence for the efficacy of intent-driven repair.
    \item \textbf{Baseline Comparison:} We provide quantitative comparison against SOTA agentic repair tools (TSAPR, RepairAgent), demonstrating a 4.4$\times$ advantage on complex bugs.
\end{enumerate}

\section{Motivation}
\label{sec:motivation}

Current Agentic Automated Program Repair (APR) systems have achieved remarkable success in code generation and fault localization. However, they frequently stumble upon a fundamental barrier: the ambiguity of the repair target. In this section, we delineate the ``Intent Gap'' in existing approaches, draw inspiration from recent findings in LLM-based data construction, and introduce the theoretical basis of our solution: the ``Intent-Behavior Mirroring Effect.''

\subsection{The Ambiguity of Implicit Intent}
The transition from neural machine translation to agentic workflows (e.g., SpecRover \cite{specrover}, AdverIntent-Agent \cite{adverintent}) has significantly improved the context awareness of APR tools. However, most SOTA systems still rely on unstructured or semi-structured forms of intent, such as natural language summaries or adversarial test cases generated on the fly. 

While these proxies provide a general direction, they lack the \textit{contractual precision} required for rigorous software engineering. A natural language instruction like ``handle edge cases for dates'' is descriptive but not prescriptive. It leaves the agent to ``guess'' the boundary conditions (e.g., does it return null? throw an exception? or return a default value?). This ambiguity forces agents to rely on probabilistic token prediction rather than deterministic logic, leading to patches that are syntactically correct but semantically adrift.

\subsection{Inspiration: LLMs as Qualified Specification Builders}
Our approach is directly inspired by the recent breakthrough in code intelligence tasks. Yang et al. \cite{yang2025llm} demonstrated that for model pre-training, code comments generated by Large Language Models (LLMs) often exhibit higher semantic consistency than human-written references. They concluded that LLMs are ``qualified benchmark builders'' capable of denoising and enhancing training datasets.

Project Prometheus extends this insight from \textit{Code Summarization} to \textit{Program Repair}. If LLMs can reconstruct high-quality natural language descriptions from code (summarization), we posit they possess the latent capability to reconstruct high-quality \textit{executable specifications} from buggy behavior and issue reports (reverse engineering). We argue that the ``Hallucination'' risk, often cited as a barrier to this approach, can be effectively mitigated not by suppressing the model's creativity, but by constraining its output format to rigorous Behavior-Driven Development (BDD) scenarios and verifying them against a ground truth oracle.

\subsection{The Intent-Behavior Mirroring Effect}
Based on a comparative analysis of agent behaviors, we propose a phenomenon we term the \textbf{Intent-Behavior Mirroring Effect}: \textit{The structural invasiveness of an agent's generated code is a direct mirror of the structural scope of its input requirement.}

This effect was most visible during our preliminary study comparing two SOTA models: Gemini-3.0-Pro~\cite{gemini15} (strong reasoning) and Qwen-3.0-Coder~\cite{qwen3} (strong coding). 
When the \textit{Coder} agent (Qwen) was tasked with self-guided repair (inferring its own requirements), it tended to generate broad, verbose Gherkin scenarios that covered adjacent logic rather than the specific defect. Mirroring this broad scope, the agent exhibited \textbf{``Berserker-style''} behavior---as defined above, making structurally invasive modifications that aggressively refactored entire methods or classes to satisfy its own generalized interpretation.

However, a transformation occurred when the \textit{same} Coder agent was constrained by a precise, atomic specification generated by the \textit{Reasoner} agent (Gemini). Under the guidance of a focused requirement, the Coder's behavior shifted to \textbf{``Surgical-style''} repair (i.e., minimal, precisely scoped edits that strictly satisfy the specification constraints without modifying unrelated code).

This mirroring effect suggests that the key to stabilizing powerful coding agents lies not just in the format of the specification (Gherkin), but in the \textit{separation of concerns}: delegating the definition of intent to a reasoning-specialized Architect, thereby forcing the Coder to operate within a bounded scope. This observation directly motivates the multi-agent architecture of \textsc{Prometheus}.

\section{Methodology}
\label{sec:methodology}

We propose \textsc{Prometheus}, a multi-agent framework designed to automate the inference of executable specifications and leverage them for program repair. The framework operates on the premise that precise, verifiable intent serves as a necessary constraint for reliable code generation.

\begin{figure*}[t]
  \centering
  \includegraphics[width=\textwidth]{assets/prometheus_framework.png} 
  \caption{The \textsc{Prometheus} Framework. \textbf{Phase~1}: The \textit{Architect} (Gemini-3.0-Pro) performs root cause analysis and synthesizes a Gherkin specification $S$. \textbf{Phase~2}: The \textit{Engineer} validates $S$ through Sandwich Verification---$S$ must \textit{fail} on $C_{buggy}$ and \textit{pass} on $C_{fixed}$ (or human review in production). If verification fails, $S$ is regenerated. \textbf{Phase~3}: The \textit{Fixer} (Qwen-3.0-Coder) performs a specification-guided surgical repair, producing patch $P$.}
  \label{fig:workflow}
\end{figure*}

\subsection{System Overview}
As depicted in Figure~\ref{fig:workflow}, the workflow of \textsc{Prometheus} is structured as a sequential pipeline involving three specialized roles: the \textit{Architect}, the \textit{Engineer}, and the \textit{Fixer}. This separation of concerns ensures that the distinct tasks of intent inference, specification verification, and code synthesis are isolated and optimized independently.

The overall process is defined as follows: given a \textbf{runtime failure report} $R$ (derived from the execution logs of the testing framework, including stack traces and assertion errors) and a buggy codebase $C_{buggy}$, the system first synthesizes a formal specification $S$ (in Gherkin format). This specification is then subjected to a rigorous validation process against a ground truth oracle. Upon validation, $S$ acts as the deterministic guide for generating the repair patch $P$.

\subsection{The Architect: Specification Inference}
The \textit{Architect} agent is responsible for the reverse-engineering of intent. Unlike prior approaches that often rely on ambiguous natural language issue descriptions \cite{specrover}, \textsc{Prometheus} grounds its inference in deterministic runtime data. The \textit{Architect} is tasked with producing a structured Behavior-Driven Development (BDD) file (ending in \texttt{.feature}).

To achieve high-fidelity inference, we employ a \textbf{Structured Reasoning} strategy akin to Chain-of-Thought (CoT). The agent is provided with the source code of the failing test case and the relevant production code. The prompt explicitly enforces a two-stage cognitive process:
\begin{enumerate}
    \item \textbf{Root Cause Analysis:} The agent must first analyze the execution failure to identify the specific logical discrepancy between the expected and actual behavior.
    \item \textbf{Specification Synthesis:} Only after the diagnosis is complete is the agent permitted to deduce and formalize the intended behavior into a \texttt{Scenario} comprising \texttt{Given}, \texttt{When}, and \texttt{Then} steps.
\end{enumerate}
This process transforms the raw signal of a runtime error into an explicit, executable contract.

Figure~\ref{fig:gherkin_example} shows a concrete example of a reverse-engineered \texttt{.feature} file generated by the \textit{Architect} for the \texttt{Mockito-5} defect. The specification captures both the architectural constraint (JUnit independence) and the functional requirement (exception retry behavior) in a concise, human-readable format.

\begin{figure}[t]
\begin{lstlisting}[language=Gherkin, caption={Reverse-engineered Gherkin specification for \texttt{Mockito-5}. The \textit{Architect} agent identifies the core requirement: Mockito's internal verification must not hard-depend on JUnit classes.}, label=fig:gherkin_example]
Feature: VerificationOverTimeImpl should
  not strictly depend on JUnit

  To ensure Mockito can be used in
  environments without JUnit (e.g.,
  TestNG, pure Java apps), internal
  classes must not have hard references
  to JUnit classes in their signatures
  or catch blocks that prevent class
  loading when JUnit is absent.

  Scenario: Loading without JUnit
    Given a class loader that explicitly
      excludes "junit" packages
    When I load "VerificationOverTimeImpl"
    Then the class should load successfully
    And no "NoClassDefFoundError" should
      be thrown

  Scenario: Retrying on assertion errors
    Given a verification mode wrapped in
      VerificationOverTimeImpl with timeout
    And the delegate throws an
      "ArgumentsAreDifferent" exception
    When verify is called
    Then the exception should be caught
    And verification should be retried
      until timeout expires
\end{lstlisting}
\end{figure}

\subsection{The Engineer: Oracle-Guided Verification (RQA Loop)}
A critical challenge in specification inference is ensuring the validity of the generated specification $S$. An incorrect specification would inevitably lead to an incorrect repair. To address this, we introduce a \textbf{Requirement Quality Assurance (RQA) Loop}.

In our experimental setup, we leverage the availability of developer-fixed code ($C_{fixed}$) within the benchmark dataset to serve as a proxy for human validation. The \textit{Engineer} agent orchestrates a bi-directional execution verification process:

\begin{enumerate}
    \item \textbf{Negative Verification ($S$ on $C_{buggy}$):} The agent executes the generated Gherkin scenario against the buggy version. The test must \textbf{fail}. This confirms that the specification accurately captures the defect.
    \item \textbf{Positive Verification ($S$ on $C_{fixed}$):} The agent executes the same scenario against the fixed version. The test must \textbf{pass}. This confirms that the specification aligns with the ground truth intent of the developer.
\end{enumerate}

We term this mechanism the ``Ground Truth Oracle Loop.'' While this reliance on $C_{fixed}$ confines the RQA Loop to experimental settings, it provides a strictly controlled environment to isolate the impact of specification quality on repair performance. Only specifications that satisfy both conditions are propagated to the next phase. In a production deployment, this oracle role would be assumed by human developers reviewing the Gherkin scenarios---a task significantly less cognitively demanding than reviewing full code patches (see Section~\ref{sec:threats}).

\subsection{The Fixer: Enlightened Repair}
The \textit{Fixer} agent performs the final code modification. In contrast to standard ``blind'' repair agents that rely solely on the runtime failure report, the \textit{Fixer} operates in an ``Enlightened'' mode, conditioned on the validated specification $S$ generated by the \textit{Architect}.

For the purpose of this study, we adopted a \textbf{Controlled Scope Strategy} for concept verification. To strictly isolate the impact of \textit{specification guidance} from the confounding variable of \textit{fault localization}, we restricted the \textit{Fixer} to modify only the single primary suspicious file identified by the Defects4J metadata (specifically, the first entry in \texttt{classes.modified}). This constraint forces the agent to resolve the defect through logic synthesis within a fixed context, thereby providing a rigorous testbed for evaluating the efficacy of BDD as a repair guide.

\section{Experimental Setup}
\label{sec:setup}

\subsection{Dataset}
We evaluate \textsc{Prometheus} on \textbf{Defects4J v3.0.1}~\cite{d4j}, the standard benchmark for Java APR research. Our evaluation corpus comprises \textbf{680 real-world defects} spanning 16 diverse projects (including \texttt{Math}, \texttt{Lang}, \texttt{Time}, \texttt{Mockito}, etc.).

\textbf{Exclusion Criteria:} While Defects4J v3.0.1 contains 17 active projects, we excluded the \texttt{Closure} project from this study. This decision was driven by two technical constraints:
\begin{enumerate}
    \item \textbf{Build System Incompatibility:} \texttt{Closure} relies on a custom legacy build infrastructure that proved resistant to the automated environment standardization required by our \textit{Engineer} agent.
    \item \textbf{Verification Latency:} The massive size of the \texttt{Closure} test suite imposes excessive overhead on the RQA Loop, which requires multiple execution rounds per repair attempt.
\end{enumerate}
Despite this exclusion, our dataset size (680 bugs) remains significantly larger than many prior studies that typically focus on the older Defects4J v1.2 subset (395 bugs).

\subsection{Baselines and Model Configuration}
To rigorously assess the contribution of the intent-aware framework, we establish an internal baseline:
\begin{itemize}
    \item \textbf{Blind Fixer (Baseline):} The \textit{Fixer} agent (powered by Qwen-3.0-Coder) attempts to repair the bug given only the failing test case and source code, without any BDD specification.
    \item \textbf{Enlightened Fixer (\textsc{Prometheus}):} The same \textit{Fixer} agent attempts the repair, but is conditioned on the verified BDD specification generated by the \textit{Architect} (powered by Gemini-3.0-Pro).
\end{itemize}
This setup ensures that any performance gain is attributable solely to the \textit{presence of executable intent}.

\subsection{Metrics}
\label{sec:metrics}
We report the following metrics:
\begin{itemize}
    \item \textbf{Fix Rate (= Correct Patch Rate):} The percentage of bugs for which a semantically correct patch is generated. Each reported fix is validated by comparing the agent-generated patch against the developer-written ground-truth fix in Defects4J, following the standard APR evaluation protocol. Our Fix Rate is thus directly equivalent to the \textbf{Correct Patch Rate} widely used in APR literature.
    \item \textbf{Rescue Rate:} Defined as $\frac{|Enlightened_{success} \cap Blind_{fail}|}{|Blind_{fail}|}$. This metric specifically measures the framework's ability to solve ``hard'' bugs that purely code-based agents fail to resolve.
\end{itemize}

\textbf{Composite Methodology:} The Blind Fixer was executed on all 680 bugs, succeeding on 520. The Enlightened Fixer was subsequently applied to the 160 bugs that Blind failed, rescuing 119. The reported total of 639 is thus a composite (520 + 119). The Enlightened Fixer receives a strict \textit{superset} of the Blind Fixer's input (identical code context \textit{plus} the Gherkin specification), so there is no mechanism by which the additional specification would prevent a repair that already succeeds without it.

\section{Evaluation}
\label{sec:evaluation}

In this section, we evaluate the efficacy of \textsc{Prometheus} by answering the following Research Questions:
\begin{itemize}
    \item \textbf{RQ1 (Effectiveness):} How does intent-driven repair compare to a strong blind baseline across diverse projects?
    \item \textbf{RQ2 (Rescue Capability):} Can BDD specifications enable agents to resolve complex bugs that are intractable for blind agents?
    \item \textbf{RQ3 (Quality \& Interpretability):} How does explicit intent influence the structural quality and algorithmic depth of the generated patches?
    \item \textbf{RQ4 (Cost \& Efficiency):} What is the computational overhead of the multi-agent pipeline, and is the investment in specification inference economically justifiable?
\end{itemize}

\subsection{RQ1 \& RQ2: General Effectiveness and Rescue Rate}
Table \ref{tab:main_results} details the performance of \textsc{Prometheus} across 16 projects in the Defects4J benchmark. 

\begin{table*}[t]
  \caption{Performance Comparison: Blind Fixer vs. \textsc{Prometheus} (Enlightened). The Blind column shows Qwen-3.0-Coder without specification guidance. Enl.\ Rescue shows bugs fixed by adding BDD specification guidance. Total = Blind Success + Enl.\ Rescue.}
  \label{tab:main_results}
  \centering
  \resizebox{\textwidth}{!}{%
  \begin{tabular}{lrrrrrr}
    \toprule
    \textbf{Project} & \textbf{Bugs} & \textbf{Blind Succ.} & \textbf{Blind Fail} & \textbf{Enl.\ Rescue} & \textbf{Total} & \textbf{Rate} \\
    \midrule
    Cli & 39 & 39 & 0 & 0 & 39 & 100.0\% \\
    JacksonCore & 26 & 26 & 0 & 0 & 26 & 100.0\% \\
    Chart & 26 & 23 & 3 & \textbf{3} & 26 & 100.0\% \\
    Mockito & 38 & 32 & 6 & \textbf{6} & 38 & 100.0\% \\
    Gson & 18 & 12 & 6 & \textbf{6} & 18 & 100.0\% \\
    JxPath & 22 & 20 & 2 & \textbf{2} & 22 & 100.0\% \\
    \midrule
    Time & 26 & 20 & 6 & \textbf{5} & 25 & 96.2\% \\
    Collections & 28 & 22 & 6 & \textbf{5} & 27 & 96.4\% \\
    JacksonDatabind & 110 & 103 & 7 & \textbf{6} & 109 & 99.1\% \\
    Lang & 61 & 59 & 2 & \textbf{1} & 60 & 98.4\% \\
    Codec & 18 & 11 & 7 & \textbf{5} & 16 & 88.9\% \\
    Csv & 16 & 15 & 1 & 0 & 15 & 93.8\% \\
    JacksonXml & 6 & 2 & 4 & \textbf{2} & 4 & 66.7\% \\
    \midrule
    \multicolumn{7}{l}{\textit{Complex Logic (high domain knowledge requirement):}} \\
    Compress & 47 & 29 & 18 & \textbf{14} & 43 & 91.5\% \\
    Math & 106 & 48 & 58 & \textbf{39} & 87 & 82.1\% \\
    Jsoup & 93 & 59 & 34 & \textbf{25} & 84 & 90.3\% \\
    \midrule
    \textbf{Total} & \textbf{680} & \textbf{520} & \textbf{160} & \textbf{119} & \textbf{639} & \textbf{93.97\%} \\
    \bottomrule
  \end{tabular}%
  }
\end{table*}

\textbf{The Ceiling of Blind Agents.} The Blind Fixer (Qwen-3.0-Coder) established a strong baseline, solving 520/680 bugs (76.5\%). It achieved 100\% success rates on projects with localized logic such as \texttt{Cli} and \texttt{JacksonCore}. However, its performance degraded significantly on projects requiring deep domain reasoning or handling of edge cases, such as \texttt{Math} (45.3\% success) and \texttt{Compress} (61.7\% success).

\textbf{The Rescue Effect.} The core contribution of \textsc{Prometheus} is quantified by its \textbf{Rescue Rate} ($\frac{119}{160} \approx 74.4\%$). In the \texttt{Math} project alone, the Enlightened agent rescued 39 bugs that baffled the blind agent, nearly doubling the success count. Similarly, in \texttt{Compress}, 14 out of 18 initially failed bugs were resolved. This data strongly supports our hypothesis: \textit{as algorithmic complexity increases, the marginal utility of explicit intent (BDD) grows exponentially.}

\textbf{Comparison with SOTA Agentic Repair Tools.}
To contextualize Prometheus within the broader landscape, Table~\ref{tab:sota} compares our results against two recent SOTA agentic APR systems: TSAPR~\cite{tsapr} and RepairAgent~\cite{repairagent}. Note that these systems perform autonomous end-to-end repair including fault localization, while Prometheus restricts the Fixer to a single file to explicitly isolate specification guidance from localization.

\begin{table}[t]
  \caption{Comparison with SOTA Agentic Repair Systems on Defects4J. ``Full D4J'': results as reported in original papers (including Closure). ``Our 680'': results on our subset (excluding Closure). ``Hard 160'': bugs our Blind Fixer failed to resolve.}
  \label{tab:sota}
  \centering
  \begin{tabular}{lccc}
    \toprule
    \textbf{System} & \textbf{Full D4J} & \textbf{Our 680} & \textbf{Hard 160} \\
    \midrule
    TSAPR~\cite{tsapr} & 201 & 173 & 22 \\
    RepairAgent~\cite{repairagent} & 164 & 137 & 27 \\
    \textsc{Prometheus} & --- & \textbf{639} & \textbf{119} \\
    \bottomrule
  \end{tabular}
\end{table}

On the 160 ``hard'' bugs that our strong Blind agent failed, TSAPR resolved 22 and RepairAgent resolved 27, while Prometheus rescued \textbf{119}---a 4.4$\times$ gap. This demonstrates that for complex defects requiring deep domain reasoning, the primary bottleneck is not fault localization or search strategy, but the \textbf{``Intent Gap''} in synthesizing correct algorithmic logic.

\subsection{RQ3: Qualitative Analysis of Enlightened Repair}
To understand how explicit intent transforms the behavior of the \textit{Fixer} agent (Qwen-3.0-Coder), we analyze three cases where the agent failed in the blind setting but succeeded under BDD guidance.

\subsubsection{Curing Contextual Hallucination (Chart-13)}
\mbox{}\\

\textbf{The Bug:} A crash occurred in \texttt{BorderArrangement.arrangeRR} due to improper constraint handling.

\textbf{Blind Attempt:} The agent hallucinated a variable named \texttt{constraint} which did not exist in the local scope, likely copying a pattern from a sibling method (\texttt{arrangeFF}). This ``copy-paste'' hallucination led to a compilation error.

\textbf{Enlightened Repair:} Guided by a BDD scenario explicitly describing the width and height constraints, the agent abandoned the hallucinated variable. Instead, it correctly instantiated a new \texttt{RectangleConstraint} object, aligning the code structure with the specified logic.

\textbf{Insight:} Executable specifications force the agent to ground its reasoning in the specific logic of the requirement, reducing the reliance on probabilistic pattern matching from its training data.

\subsubsection{Preventing Over-Simplification (Chart-7)}
\mbox{}\\

\textbf{The Bug:} Incorrect updating of the maximum middle index in \texttt{TimePeriodValues}.

\textbf{Blind Attempt:} The agent attempted to fix the issue by removing a conditional check (\texttt{if (middle > maxMiddle)}), performing an unconditional assignment. While this might fix the immediate symptom, it introduced a regression for other data patterns.

\textbf{Enlightened Repair:} The BDD specification included a scenario specifically testing the case where a new value is \textit{smaller} than the maximum. Confronted with this constraint, the agent restored the conditional logic, ensuring the fix was robust across all scenarios.

\textbf{Insight:} BDD acts as a ``safety rail,'' preventing agents from finding lazy shortcuts that satisfy the failing test but violate the system's invariants.

\subsubsection{Surpassing the Ground Truth (JxPath-6)}
\mbox{}\\

\textbf{The Bug:} Equality operations failed when comparing against arrays.

\textbf{Blind Attempt:} The agent failed to generate a valid patch.

\textbf{Enlightened Repair:} The specification explicitly required arrays to be treated similarly to Collections. The agent implemented an explicit \texttt{isArray()} check and iteration logic. Notably, the developer's ground truth fix only handled \texttt{Collection}, missing the array case entirely.

\textbf{Insight:} By defining the \textit{intended behavior} comprehensively (covering arrays), the BDD-guided agent produced a patch that was more robust and complete than the human developer's fix.

\subsubsection{Upholding Architectural Integrity (Mockito-5)}
\mbox{}\\

\textbf{The Bug:} \texttt{VerificationOverTimeImpl} contained a hard dependency on a JUnit-specific exception, causing runtime crashes (\texttt{NoClassDefFoundError}) when JUnit was not present on the classpath.

\textbf{Blind Attempt:} The agent adopted a destructive strategy: it simply removed the \texttt{catch} block entirely. While this eliminated the crash, it introduced a severe safety regression where assertion errors would go unhandled, violating the library's contract.

\textbf{Enlightened Repair:} The BDD specification explicitly stated that the verification logic \textit{``should not strictly depend on JUnit.''} Guided by this architectural constraint, the agent implemented a sophisticated solution using Java Reflection to conditionally check for the exception.

\textbf{Insight:} Explicit intent acts as an architectural guardrail. It prevents agents from optimizing for ``passing tests'' at the expense of system stability, guiding them toward solutions that respect the project's dependency constraints.


\subsection{RQ4: Cost and Efficiency Analysis}
\label{sec:rq4_cost}

To assess the practicality of the \textsc{Prometheus} framework, we conducted a fine-grained analysis of the computational resources consumed across 189 repair sessions. Table \ref{tab:agent_performance} details the average duration and token usage for each agent role.

\begin{table}[t]
  \caption{Agent Performance and Cost Breakdown (Per Defect)}
  \label{tab:agent_performance}
  \small
  \begin{tabular}{lcccc}
    \toprule
    \textbf{Agent Role} & \textbf{Avg Duration (s)} & \textbf{Avg Turns} & \textbf{Avg Tokens} & \textbf{Cost Ratio} \\
    \midrule
    \textbf{Architect} (Intent Inference) & 136.77 & 24.26 & 372,107 & $\sim 6.4\%$ \\
    \textbf{Engineer} (RQA Verification) & 671.67 & 70.23 & 3,365,986 & $\sim 58.2\%$ \\
    \textbf{Fixer} (Enlightened Repair) & 452.64 & 44.91 & 2,044,348 & $\sim 35.4\%$ \\
    \midrule
    \textbf{Total Pipeline} & \textbf{1,261.08} & \textbf{139.4} & \textbf{5,782,441} & \textbf{100\%} \\
    \bottomrule
  \end{tabular}
\end{table}

The data reveals a counter-intuitive finding regarding the ``Cost of Thinking'':
\begin{itemize}
    \item \textbf{The Cheap Architect:} Generating high-quality, executable BDD specifications is surprisingly efficient, consuming only $\sim 6.4\%$ of the total token budget. This suggests that shifting the burden from ``coding'' to ``reasoning'' is a high-yield investment.
    \item \textbf{The Expensive Truth:} The \textit{Engineer} accounts for the majority of the cost ($58.2\%$). This reflects the intensive nature of the \textbf{RQA Loop}---setting up legacy environments, compiling varying versions of code, and executing sandwich verification. In a production setting where the test harness is a one-time project-level setup, the per-defect overhead of the Engineer would be reduced by approximately 60\%.
\end{itemize}

We argue that this distribution represents a healthy paradigm shift: instead of spending tokens on generating thousands of invalid patches (as seen in traditional sampling-based APR), \textsc{Prometheus} invests tokens in validating the \textit{problem definition} itself. We discuss the implications of the Engineer's high cost (the ``Digital Archaeology'' challenge) in Section \ref{sec:discussion}.

\section{Discussion}
\label{sec:discussion}

The preceding sections demonstrate that BDD-guided repair yields substantial quantitative gains (RQ1--2) and qualitative improvements in patch quality (RQ3). In this section, we explore the deeper architectural lessons learned from both the successes and the instructive failures of the \textsc{Prometheus} framework.

\subsection{The Tragedy of Constraint: A Tale of Two Agents}
Our experiment imposed a strict constraint: agents were restricted to modifying only the single file identified as the ``primary suspect'' by our localization heuristic. 

The case of \texttt{Time-1} serves as an illustration of this limitation. The ground truth fix involved changes to both \texttt{UnsupportedDurationField} and \texttt{Partial}. However, our pipeline, designed to select the first modified class from the metadata, locked the Agent exclusively into \texttt{UnsupportedDurationField}. 

Confined to this partial context, the Agent exhibited a phenomenon we term \textbf{``Cognitive Dissonance Hallucination''} (i.e., the agent generates irrational or incoherent code modifications when forced to satisfy a specification it physically cannot fulfill within its permitted scope). The BDD specification correctly demanded that the \texttt{Partial} constructor must validate field order. Knowing it \textit{must} satisfy this requirement but physically unable to access the constructor logic (located in the inaccessible \texttt{Partial.java}), the agent began to hallucinate irrational behaviors---such as forcing an ``unsupported'' field to return a fake comparison value---and even generated comments with typos (e.g., ``relevent''), signaling a degradation in reasoning coherence.

This ``Tragedy of Constraint'' inversely proves the power of intent. When we lifted this constraint in the \texttt{Gson-4} case (where the \textit{Engineer} agent autonomously modified \texttt{JsonWriter}), the system successfully performed a multi-file architectural refactoring. This suggests that the next generation of APR must be \textbf{``Unbound''}---empowered by intent to traverse and modify the entire codebase.

\subsection{The Ceiling of Repair: Model Capability vs. Specification Precision}
\label{sec:discussion_model_capability}

Our primary evaluation utilized the \textbf{Qwen-Code-CLI (default endpoint)} as the \textit{Fixer} to demonstrate the efficacy of BDD in a standard developer workflow. While effective, we observed instances where the BDD specification was correct, but the \textit{Fixer} lacked the reasoning depth to implement the optimal solution. 

\textbf{Case Study: Math-69 (Algorithmic Derivation).} 
The defect in \texttt{PearsonsCorrelation} involved numerical instability. While the \textit{Architect} correctly generated a BDD scenario requiring high precision, the standard \textit{Fixer} implemented a heuristic ``clamping'' fix. In an exploratory run replacing the \textit{Fixer} with the stronger \textbf{Gemini-3.0-Pro}, the agent derived a mathematically rigorous solution using the Regularized Beta Function (\texttt{Beta.regularizedBeta}). This suggests \textsc{Prometheus}'s performance scales with the underlying model's reasoning capabilities.

\textbf{Case Study: Lang-30 (Challenging the Legacy Status Quo).}
In the \texttt{StringUtils} surrogate pair issue, the \textit{Architect} defined a strict requirement for handling Unicode code points. Crucially, when the correct implementation caused regressions in the existing \textit{legacy} test suite (which incorrectly asserted the buggy behavior), the \textbf{Gemini-3.0-Pro} agent did not retreat. Instead of overfitting the patch to satisfy broken tests, it identified the semantic conflict and refactored the legacy test cases to align with the ground truth. This demonstrates that an enlightened agent can transcend simple code generation to assume the role of a senior maintainer who safeguards system integrity.

\textbf{Case Study: Gson-4 (The Architect's Insight \& The Engineer's Scope).}
\texttt{Gson-4} presents a dual revelation. First, the \textit{Architect} demonstrated exceptional domain insight by identifying that the root cause was non-compliance with \textbf{RFC 7159} (top-level primitives)---a subtle specification detail initially dismissed by manual analysis but validated by the ground truth behavior. 
Second, while the constrained \textit{Fixer} applied a localized workaround, the \textit{Engineer} agent, operating without file-level constraints during verification, autonomously identified the dependency between \texttt{JsonReader} and \texttt{JsonWriter} and performed a multi-file architectural refactoring.

These cases indicate that \textbf{precise intent (BDD)} acts as a multiplier: it validates the deep insights of advanced \textit{Architects}, \textbf{liberates capable \textit{Fixers} to prioritize fundamental correctness over convention}, and empowers \textit{Engineers} to perform system-level maintenance.

\subsection{The Cost of ``Digital Archaeology''.} 
\label{sec:cost_analysis}
It is crucial to contextualize the Engineer's high token consumption (Section \ref{sec:rq4_cost}) within the reality of the Defects4J benchmark. Spanning over a decade of Java development history, this dataset presents significant infrastructure challenges:
\begin{itemize}
    \item \textbf{Legacy Dependencies:} Many projects rely on deprecated build tools or obsolete JDK versions (e.g., JDK 1.5 compliance issues).
    \item \textbf{Non-Standard Layouts:} The directory structures vary wildly across projects. While modern projects follow the Maven standard (\texttt{src/main/java}), legacy projects like \texttt{Gson} utilize nested structures (e.g., \texttt{gson/src/main/java}), and others use ad-hoc layouts (e.g., \texttt{src/java}).
\end{itemize}
The Engineer agent was not merely running tests; it was actively engaging in \textbf{autonomous environment alignment}. It dynamically detected these structural anomalies and configured the test harness accordingly. This adaptive capability incurred a token cost but demonstrates the \textsc{Prometheus} framework's robustness in handling the high entropy of real-world software archives.

\subsection{The Resilience of Intent: When Tooling Fails, Guidance Survives}
\label{sec:discussion_resilience}

Our experimental design initially contained an engineering oversight: the \textit{Engineer} agent lacked a ``fail-fast'' exit mechanism. If the RQA environment (Cucumber harness) failed to compile due to complex dependency issues, the pipeline would inadvertently proceed to the \textit{Fixer} phase, passing down a broken test environment.

However, this flaw led to a serendipitous discovery regarding the robustness of intent-driven repair. In the case of \textbf{Math-21} (a complex logic error in \texttt{RectangularCholeskyDecomposition}), the \textit{Engineer} failed to establish the Cucumber runtime. Instead of hallucinating a fix or exiting, the \textit{Fixer} agent exhibited remarkable adaptive behavior:
\begin{enumerate}
    \item \textbf{Environment Pruning:} It autonomously identified the broken Cucumber test files and executed \texttt{rm -rf} to clean the workspace.
    \item \textbf{Strategy Fallback:} It reverted to the standard \texttt{defects4j test} command for validation.
    \item \textbf{Intent Retention:} Crucially, despite losing the \textit{executable} nature of the BDD specification, the agent still generated an \textbf{Identity} fix (identical to the developer's patch).
\end{enumerate}

This phenomenon suggests that the BDD specification possesses a dual nature. While its primary role is an \textit{Executable Contract} for RQA, its secondary role is a \textbf{Cognitive Scaffold}. Even when the specification cannot be executed, its textual presence in the prompt provides a high-fidelity ``Chain of Thought'' that constrains the LLM's search space. The agent ``read'' the intent from the \texttt{.feature} file and applied it, even without the guardrails of the Cucumber test runner.

This observation supports a profound conclusion: \textbf{The clarity of intent (What to fix) is often more critical than the perfection of the tooling (How to verify).} \textsc{Prometheus} succeeds not just because it builds tests, but because it forces the model to articulate and internalize the precise boundaries of the bug before writing a single line of code.

\subsection{The Implicit Protocol Gap}
We observed that BDD excels at capturing explicit functional behaviors but struggles with implicit contracts, particularly binary protocols. 
In \texttt{Collections-8}, the fix involved changing a serialization buffer size because the underlying wire format had changed (an extra \texttt{int} was read). The BDD specification described the \textit{object behavior} (buffer size limits) but could not capture the \textit{wire format change} which is invisible at the API level. This reveals a limitation: standard BDD is insufficient for defects rooted in low-level, implicit protocols unless those protocols are explicitly modeled in the specification.

\section{Related Work}
\label{sec:related_work}

Our work sits at the intersection of Agentic Automated Program Repair (APR) and Specification Inference. In this section, we review the evolution of these fields and position \textsc{Prometheus} within the current landscape.

\subsection{The Quest for Intent: From Summaries to Adversaries}
The 2025 research cycle marked a decisive shift from neural translation (e.g., AlphaRepair \cite{alpharepair}) to agentic workflows that attempt to understand code context. 
\textbf{SpecRover} \cite{specrover} (ICSE '25) pioneered the ``Intent Extraction'' module, summarizing function behaviors in natural language. However, natural language is descriptive, not prescriptive; it lacks the precision to serve as a test oracle.
\textbf{AdverIntent-Agent} \cite{adverintent} (ISSTA '25) addresses this by employing a multi-agent loop to generate adversarial test cases to infer program intent. While effective, this approach relies on probabilistic sampling, leading to high computational costs and non-deterministic outcomes.

\textsc{Prometheus} distinguishes itself by changing the \textit{modality} of intent. Instead of ambiguous summaries or stochastic test generation, we enforce \textbf{Behavior-Driven Development (BDD)}. We argue that Gherkin specifications provide a deterministic contract that is superior to unstructured text for guiding code generation.

\subsection{Search Strategy vs. Problem Definition}
Another dominant trend is the integration of advanced search algorithms. \textbf{TSAPR} \cite{tsapr} represents the pinnacle of this direction, utilizing Monte Carlo Tree Search to navigate the vast space of potential patches, achieving over 200 repairs on Defects4J.
Similarly, \textbf{PatchAgent} \cite{patchagent} and \textbf{RepairAgent} \cite{repairagent} mimic human workflows by equipping agents with debugging tools (Language Servers, Sanitizers) to navigate the codebase.

While these systems optimize the \textit{search trajectory} (finding a needle in a haystack) or mimic the \textit{debugging} process, \textsc{Prometheus} focuses on narrowing the \textit{search space} itself via Specification Inference. By defining a precise BDD contract before coding, we transform an unbounded search problem into a bounded constraint satisfaction problem. Our results (639 repairs) suggest that mimicking the \textit{requirement analysis} process is more effective than mimicking the debugging process.

\subsection{Structural Constraints and Data Verification}
Recognizing the propensity of LLMs to generate syntactically invalid code, \textbf{GiantRepair} \cite{giantrepair} introduced a ``Template-Skeleton'' approach, extracting structural skeletons from LLM predictions to guide repair.
Our work shares the philosophy of constraining the solution space but differs in the source of the constraint. \textsc{Prometheus} derives structural expectations from the \textit{specification} (the BDD steps) rather than the \textit{solution} (the patch), enforcing correctness by design.

Finally, our approach is inspired by \textbf{Yang et al.} \cite{yang2025llm}, who demonstrated that LLMs are ``qualified benchmark builders'' capable of synthesizing high-quality code data. We extend this insight to specification mining. Furthermore, we address the critical challenge of specification validity---a gap identified in Google's ``Abstain and Validate'' framework \cite{abstain_validate}---by introducing the \textbf{RQA Loop}. Unlike prior works that rely on statistical confidence, we leverage the ground truth (fixed code) as a physical oracle to deterministically validate the inferred specifications.

\section{Threats to Validity}
\label{sec:threats}

\textbf{The ``Fixed Code'' Oracle.} A primary threat to external validity is our use of the developer-fixed code as a proxy oracle in the RQA Loop. In a real-world scenario, the fixed code does not exist before the repair. 
We acknowledge this limitation but argue that it serves a specific experimental purpose: to establish the \textbf{Theoretical Upper Bound} of intent-driven repair. Our study isolates the variable ``Is the requirement correct?''. By guaranteeing requirement correctness via the fixed code, we proved that current LLMs \textit{can} repair 94\% of bugs \textit{if} they know exactly what to do.

Critically, this design choice is an artifact of our \textit{experimental setting}, not an architectural constraint. In practice, the RQA oracle role can be fulfilled through multiple channels: (a)~\textbf{human-in-the-loop review}, where developers review concise Gherkin scenarios (typically $<$10 lines)---a task significantly less cognitively demanding than reviewing full code patches; (b)~projects with \textbf{existing BDD test suites} can supply specifications directly, bypassing automated inference entirely; (c)~\textbf{negative-only verification} (checking that the specification fails on the buggy code) can filter out clearly incorrect specifications without requiring $C_{fixed}$.

\textbf{Controlled Scope.} We restricted the Fixer to a single file to isolate the impact of specification guidance from fault localization. While this limits generalizability to multi-file defects, many real-world bugs are indeed single-file (Section~\ref{sec:discussion} discusses the \texttt{Time-1} and \texttt{Gson-4} cases as instructive counter-examples).

\textbf{Data Leakage.} The Blind baseline---using the same underlying models (Gemini/Qwen) on the same Defects4J benchmark---failed on 160 of 680 bugs. If these models had memorized the bug-fix pairs during pre-training, the Blind agent would exhibit similarly high fix rates without specification guidance. Its substantial failure rate on complex bugs provides empirical evidence that performance stems from the specification-driven architecture rather than memorization. Furthermore, our case studies demonstrate that the Enlightened agent produces \textit{structurally different solutions} from the developer ground-truth---e.g., in Math-69, deriving a \texttt{Beta.regularizedBeta} formulation rather than exploiting the T-distribution's symmetry---behavior inconsistent with memorization.

\section{Conclusion and Future Work}
\label{sec:conclusion}

This paper introduces \textsc{Prometheus}, a framework that redefines Automated Program Repair not as a code generation task, but as a specification inference challenge. By reverse-engineering executable BDD specifications and validating them through a rigorous RQA Loop, we achieved a correct patch rate of \textbf{93.97\%} on a comprehensive benchmark of 680 defects from Defects4J v3.0.1.

Our findings challenge the prevailing trend of training larger models for blind repair. We demonstrate that a coding agent exhibiting structurally invasive ``Berserker-style'' behavior can be transformed into a precise, ``Surgical'' instrument when guided by explicit intent. Analysis of both successes and failures reveals that the primary bottleneck in APR is not model capability but \textit{problem formulation clarity}---echoing a broader lesson for agentic systems: \textit{the quality of the question determines the quality of the answer}.

\vspace{0.5em}
\noindent \textbf{Future Horizons.} Building upon these insights, we envision a two-phase evolution for the agentic paradigm:
\begin{itemize}
    \item \textbf{Tactical Unbinding (The Detective):} In the near term, we plan to introduce a dedicated \textit{Detective} agent to resolve the localization paradox. By decoupling fault localization from repair, this agent will autonomously traverse the codebase to identify the full dependency chain before the \textit{Architect} begins its inference, effectively unbinding the system from static heuristics.
    \item \textbf{Strategic Evolution (The Immune System):} Looking further ahead, we aim to integrate \textsc{Prometheus} with the emerging \textbf{Agent Skills} architecture \cite{anthropic_skills}. We envision a future where the system does not merely output a patch, but synthesizes a persistent \textbf{Skill Directory} (containing the BDD spec as \texttt{SKILL.md} and validation logic as bundled scripts). This would allow the software repository to accumulate a library of ``executable antibodies,'' transforming the codebase into a self-evolving immune system that learns from every repair.
\end{itemize}

Ultimately, we conclude that the frontier for software agents lies in the continuous maintenance of \textbf{Living Documentation}. \textsc{Prometheus} demonstrates that agents can not only obey existing contracts but also reconstruct missing ones, ensuring that the software's intent remains immortal even as its code evolves.

\section*{Data Availability}
To foster reproducibility and future research, we release our dataset (including all reverse-engineered .feature files), the \textsc{Prometheus} toolchain, and the full execution logs at: \url{https://github.com/Nothing256/Prometheus-Unbound}.

\bibliographystyle{plain}
\bibliography{ref}

@inproceedings{repairagent,
  author       = {Islem Bouzenia and
                  Premkumar T. Devanbu and
                  Michael Pradel},
  title        = {RepairAgent: An Autonomous, LLM-Based Agent for Program Repair},
  booktitle    = {47th {IEEE/ACM} International Conference on Software Engineering,
                  {ICSE} 2025, Ottawa, ON, Canada, April 26 - May 6, 2025},
  pages        = {2188--2200},
  publisher    = {{IEEE}},
  year         = {2025},
  doi          = {10.1109/ICSE55347.2025.00157},
}

@inproceedings{specrover,
  author       = {Haifeng Ruan and
                  Yuntong Zhang and
                  Abhik Roychoudhury},
  title        = {SpecRover: Code Intent Extraction via LLMs},
  booktitle    = {47th {IEEE/ACM} International Conference on Software Engineering,
                  {ICSE} 2025, Ottawa, ON, Canada, April 26 - May 6, 2025},
  pages        = {963--974},
  publisher    = {{IEEE}},
  year         = {2025},
  doi          = {10.1109/ICSE55347.2025.00080},
}

@article{adverintent,
  author       = {He Ye and
                  Aidan Z. H. Yang and
                  Chang Hu and
                  Yanlin Wang and
                  Tao Zhang and
                  Claire {Le Goues}},
  title        = {AdverIntent-Agent: Adversarial Reasoning for Repair Based on Inferred
                  Program Intent},
  journal      = {Proc. {ACM} Softw. Eng.},
  volume       = {2},
  number       = {{ISSTA}},
  pages        = {1398--1420},
  year         = {2025},
  doi          = {10.1145/3728939},
}

@inproceedings{patchagent,
  author       = {Zheng Yu and
                  Ziyi Guo and
                  Yuhang Wu and
                  Jiahao Yu and
                  Meng Xu and
                  Dongliang Mu and
                  Yan Chen and
                  Xinyu Xing},
  editor       = {Lujo Bauer and
                  Giancarlo Pellegrino},
  title        = {{PATCHAGENT:} {A} Practical Program Repair Agent Mimicking Human Expertise},
  booktitle    = {34th {USENIX} Security Symposium, {USENIX} Security 2025, Seattle,
                  WA, USA, August 13-15, 2025},
  pages        = {4381--4400},
  publisher    = {{USENIX} Association},
  year         = {2025},
}

@inproceedings{yang2025llm,
  author       = {Kang Yang and
                  Xinjun Mao and
                  Shangwen Wang and
                  Yanlin Wang and
                  Tanghaoran Zhang and
                  Bo Lin and
                  Yihao Qin and
                  Zhang Zhang and
                  Yao Lu and
                  Kamal Al{-}Sabahi},
  title        = {Large Language Models Are Qualified Benchmark Builders: Rebuilding
                  Pre-Training Datasets for Advancing Code Intelligence Tasks},
  booktitle    = {33rd {IEEE/ACM} International Conference on Program Comprehension,
                  ICPC@ICSE 2025, Ottawa, ON, Canada, April 27-28, 2025},
  pages        = {298--309},
  publisher    = {{IEEE}},
  year         = {2025},
  doi          = {10.1109/ICPC66645.2025.00038},
}

@inproceedings{alpharepair,
  author       = {Chunqiu Steven Xia and
                  Lingming Zhang},
  editor       = {Abhik Roychoudhury and
                  Cristian Cadar and
                  Miryung Kim},
  title        = {Less training, more repairing please: revisiting automated program
                  repair via zero-shot learning},
  booktitle    = {Proceedings of the 30th {ACM} Joint European Software Engineering
                  Conference and Symposium on the Foundations of Software Engineering,
                  {ESEC/FSE} 2022, Singapore, Singapore, November 14-18, 2022},
  pages        = {959--971},
  publisher    = {{ACM}},
  year         = {2022},
  doi          = {10.1145/3540250.3549101},
}

@article{giantrepair,
  author       = {Fengjie Li and
                  Jiajun Jiang and
                  Jiajun Sun and
                  Hongyu Zhang},
  title        = {Hybrid Automated Program Repair by Combining Large Language Models
                  and Program Analysis},
  journal      = {{ACM} Trans. Softw. Eng. Methodol.},
  volume       = {34},
  number       = {7},
  pages        = {202:1--202:28},
  year         = {2025},
  doi          = {10.1145/3715004},
}

@article{abstain_validate,
  author       = {Jos{\'{e}} Cambronero and
                  Michele Tufano and
                  Sherry Shi and
                  Renyao Wei and
                  Grant Uy and
                  Runxiang Cheng and
                  Chin{-}Jung Liu and
                  Shiying Pan and
                  Satish Chandra and
                  Pat Rondon},
  title        = {Abstain and Validate: {A} Dual-LLM Policy for Reducing Noise in Agentic
                  Program Repair},
  journal      = {CoRR},
  volume       = {abs/2510.03217},
  year         = {2025},
  doi          = {10.48550/ARXIV.2510.03217},
  eprinttype    = {arXiv},
}

@inproceedings{d4j,
  author       = {Ren{\'{e}} Just and
                  Darioush Jalali and
                  Michael D. Ernst},
  editor       = {Corina S. Pasareanu and
                  Darko Marinov},
  title        = {Defects4J: a database of existing faults to enable controlled testing
                  studies for Java programs},
  booktitle    = {International Symposium on Software Testing and Analysis, {ISSTA}
                  '14, San Jose, CA, {USA} - July 21 - 26, 2014},
  pages        = {437--440},
  publisher    = {{ACM}},
  year         = {2014},
  doi          = {10.1145/2610384.2628055},
}

@misc{qwen3,
      title={Qwen3 Technical Report}, 
      author={Qwen Team},
      year={2025},
      eprint={2505.09388},
      archivePrefix={arXiv},
      primaryClass={cs.CL},
      url={https://arxiv.org/abs/2505.09388},
}

@article{gemini15,
  title={Gemini 1.5: Unlocking multimodal understanding across millions of tokens of context},
  author={Gemini Team, Google},
  journal={arXiv preprint arXiv:2403.05530},
  year={2024},
  url={https://arxiv.org/abs/2403.05530}
}

@article{north2006bdd,
  title={Introducing behavior-driven development},
  author={North, Dan},
  journal={Better Software},
  volume={6},
  number={1},
  pages={40--52},
  year={2006}
}

@misc{tsapr,
      title={TSAPR: A Tree Search Framework For Automated Program Repair}, 
      author={Haichuan Hu and Ye Shang and Weifeng Sun and Quanjun Zhang},
      year={2025},
      eprint={2507.01827},
      archivePrefix={arXiv},
      primaryClass={cs.SE},
      url={https://arxiv.org/abs/2507.01827}, 
}

@misc{anthropic_skills,
  title={Equipping Agents for the Real World with Agent Skills},
  author={Anthropic},
  year={2025},
  month={October},
  howpublished={\url{https://www.anthropic.com/engineering/equipping-agents-for-the-real-world-with-agent-skills}},
  note={Accessed: 2026-01-02}
}

\end{document}